\documentclass[fleqn,usenatbib]{mnras}

\usepackage{newtxtext,newtxmath}

\usepackage[T1]{fontenc}
\usepackage{tablefootnote}
\DeclareRobustCommand{\VAN}[3]{#2}
\let\VANthebibliography\thebibliography
\def\thebibliography{\DeclareRobustCommand{\VAN}[3]{##3}\VANthebibliography}

\usepackage{graphicx}	
\usepackage{amsmath}	

\usepackage{xcolor}
\usepackage{siunitx}

\newcommand{\fig}[1]{Figure~\ref{fig:#1}}
\renewcommand{\sec}[1]{Section~\ref{sec:#1}}
\newcommand{\tab}[1]{Table~\ref{tab:#1}}

\newcommand{\mumetre}{\SI{}{\micro\meter}}
\newcommand{\simba}{\textsc{Simba}}

\title[Orientation bias]{An Orientation Bias in Observations of Submillimetre Galaxies}

\author[C. C. Lovell et al.]{C. C. Lovell,$^{1}$\thanks{E-mail: c.lovell@herts.ac.uk (CCL)}
J. E. Geach,$^{1}$
R. Dav\'{e},$^{2,3,4}$
D. Narayanan$^{5,6,7}$,
K.~E.~K. Coppin$^{1}$,
Q. Li$^{5}$, \newauthor
M. Franco$^{1}$ \&
G. C. Privon$^{8}$
\\
$^{1}$Centre for Astrophysics Research, Department of Physics, Astronomy \& Mathematics, University of Hertfordshire, Hatfield AL10 9AB\\
$^{2}$Institute for Astronomy, Royal Observatory, University of Edinburgh, Edinburgh EH9 3HJ\\
$^{3}$University of the Western Cape, Bellville, Cape Town 7535, South Africa\\
$^{4}$South African Astronomical Observatories, Observatory, Cape Town 7925, South Africa\\
$^{5}$Department of Astronomy, University of Florida, 211 Bryant Space Sciences Center, Gainesville, FL, USA\\
$^{6}$University of Florida Informatics Institute, 432 Newell Drive, CISE Bldg E251, Gainesville, FL, USA\\
$^{7}$Cosmic Dawn Center, Niels Bohr Institute, University of Copenhagen and DTU-Space, Technical University of Denmark\\
$^{8}$National Radio Astronomy Observatory, 520 Edgemont Rd., Charlottesville, VA, 22903, USA
}

\pubyear{2022}

\begin{document}
\label{firstpage}
\pagerange{\pageref{firstpage}--\pageref{lastpage}}
\maketitle

\begin{abstract}
Recent high-resolution interferometric images of submillimetre galaxies (SMGs) reveal fascinatingly complex morphologies.
This raises a number of questions: how does the relative orientation of a galaxy affect its observed submillimetre emission, and does this result in an `orientation bias' in the selection and analysis of such galaxies in flux-limited cosmological surveys?
We investigated these questions using the \textsc{Simba} cosmological simulation paired with the dust radiative transfer code \textsc{Powderday}.
We selected eight simulated SMGs ($S_{850}\gtrsim2$ mJy) at $z = 2$, and measured the variance of their `observed' emission over 50 random orientations.
Each galaxy exhibits significant scatter in its emission close to the peak of the thermal dust emission, with variation in flux density of up to a factor of 2.7.
This results in an appreciable dispersion in the inferred dust temperatures and infrared luminosities ($16^{\mathrm{th}}-84^{\mathrm{th}}$ percentile ranges of 5\,K and 0.1\,dex, respectively) and therefore a fundamental uncertainty in derived parameters such as dust mass and star formation rate ($\sim$30\% for the latter using simple calibrations).
Using a Monte Carlo simulation we also assessed the impact of orientation on flux-limited surveys, finding a bias in the selection of SMGs towards those with face--on orientations, as well as those at lower redshifts.
We predict that the orientation bias will affect flux-limited single-dish surveys, most significantly at THz frequencies, and this bias should be taken into account when placing the results of targeted follow--up studies in a statistical context.
\end{abstract}




\section{Introduction}

Submillimetre astronomers must navigate a morass of telluric absorption to get a clear view of the cosmos. Nature allows terrestrial observers but a few glimpses through the submillimetre windows, primarily at approximately 350, 450 and 850\mumetre. Even so, like children stretching up to peer over the sill, Earth-bound observers must strain as close as possible to the stars, seeking high and dry sites to minimise the deleterious column of water vapour, the bane of submillimetre observers.
Above the atmosphere the submillimetre perspective is in principle clear, but at the cost of resolution: space telescopes such as {\it Herschel} \citep{pilbratt_herschel_2010} cannot yet rival the scale of their ground-based counterparts, so while clear, the view is confused.

In deep extragalactic ground-based surveys conducted through one of the submillimetre windows, or at any given wavelength for that matter, we refer to, for example, `350\mumetre-selected' or `850\mumetre-selected' submillimetre galaxies (SMGs).
Regardless of their selection, SMGs broadly represent a class of rapidly star-forming galaxies at high redshift \citep[$z>1$;][see \cite{casey_dusty_2014} for a review]{smail_deep_1997,barger_submillimetre-wavelength_1998,hughes_high-redshift_1998,lilly_canada-united_1999,chapman_redshift_2005}.
Their spectral energy distribution (SED) is dominated by thermal emission from carbonaceous and silicate particles -- dust -- heated by the interstellar radiation field \citep{hildebrand_determination_1983}.
Thus, SMGs' far-infrared SEDs approximately follow blackbody emission, peaking at a rest-frame wavelength of 70--125\mumetre{} \citep{casey_dusty_2014}.
For typical dust temperatures of a few tens of Kelvin, the bulk of the far-infrared SED is redshifted to submillimetre/millimetre wavelengths for sources at cosmological distances, and  with a negative {\it K}-correction in the Rayleigh-Jeans regime, galaxies have roughly constant (or even increasing) brightness at submillimetre wavelengths over $z \approx 1\text{--}10$ for a given luminosity \citep{blain_submillimeter_2002}.


Although submillimetre surveys provide, arguably, an `ideal' census of infrared-luminous (and therefore dust-obscured) activity over cosmic time, it is critical to assess the selection biases inherent in single band cosmological surveys.
For instance, there is a well-known flux density dependence on the intrinsic dust temperature of galaxies stemming from Wien's law which should be included in any assessment of survey depth or completeness, and which is most prominent for wavelengths beyond 500 \mumetre\ \citep[e.g.~][]{blain_galaxygalaxy_1996,eales_canada-uk_2000,blain_accurate_2004,chapman_population_2004,casey_confirming_2009}.
As an example, consider an SMG  at $z=2$; a change in peak dust temperature of $\Delta T = 20$\,K corresponds to a 850\mumetre--1.2mm flux density variation of up to 1\,dex \citep{casey_dusty_2014}.
Thus, a flux-limited survey will tend to result in a temperature bias in the selection of galaxies at a given redshift.
The fact that the redshift causes the thermal peak to move through the submillimetre windows could also result in a bias in the redshift distribution of SMGs selected at a fixed wavelength, such that the brightest galaxies selected at long wavelengths are typically at higher redshift \citep{smolcic_millimeter_2012,simpson_scuba-2_2017,brisbin_alma_2017,miettinen_alma_2017,dudzeviciute_alma_2020}.

There have long been hints, from the few bright \citep{hodge_evidence_2012} or highly lensed sources \citep{swinbank_intense_2010} available, that the detailed distribution of submillimetre emission in individual SMGs is likely to be complex, but for many years actually measuring the submillimetre morphology was out of reach due to the coarse resolution of single dish facilities and the relatively low sensitivity of interferometric arrays.
Substantial progress has been made in the past decade, starting with the Plateau de Bure Interferometer \citep[now the NOrthern Extended Millimeter Array; e.g.,][]{smolcic_millimeter_2012} and the Submillimetre Array \citep[e.g.,][]{barger_precise_2012},
and a step change with the Atacama Large Millimetre/submillimetre Array (ALMA) which has revolutionized our view of the submillimetre Universe \citep{hodge_high-redshift_2020}.
The sensitivity and resolution of ALMA is revealing the morphology of SMGs with increasing clarity \citep[e.g.,][]{carilli_anatomy_2013, hodge_kiloparsec-scale_2016, hodge_alma_2019, oteo_witnessing_2016, oteo_almacal_2017, gullberg_dust_2018, diaz-santos_multiple_2018, gullberg_alma_2019, rujopakarn_alma_2019, litke_spatially_2019}.

Here we consider the impact of the projected orientation of SMGs on their observed submillimetre emission, and investigate potential selection and analysis biases that might arise as a result.
For example, for submillimetre morphologies with axisymmetric components, the projected orientation may affect selection if the emission is optically thick, and the exact distribution of the structures responsible for the emission will determine the severity of the effect.
Moreover, such an \textit{orientation bias} will be wavelength dependent, compounding or competing with the known selection effects described above.

We take a theoretical approach, making use of the \textsc{Simba} cosmological hydrodynamic simulation \citep{dave_simba:_2019} and a sophisticated treatment for dust radiative transfer \citep{narayanan_powderday_2021} to produce multiple views of a given SMG for coeval observers at different locations.
One may question the fidelity of the simulation in producing realistic SMGs -- a concern we address -- however we make predictions that could be tested observationally, particularly with future single-dish observatories \citep{kawabe_new_2016,klaassen_atacama_2020}, which would confirm or refute an orientation bias in the selection of SMGs. Our investigation is structured as follows.
In \sec{sims} we describe the \simba\ simulations, our radiative transfer modelling, and the sample selection.
\sec{analysis} contains our analysis, including an exploration of the physical cause of the orientation dependence (\sec{explain}), the uncertainty incurred in measurements of physical parameters (\sec{uncertain}), and the bias introduced in flux-limited SMG surveys (\sec{bias}).
We discuss our results and reiterate our conclusions in \sec{conc}.
Throughout we assume a \cite{planck_collaboration_planck_2016} cosmology, and a \cite{chabrier_galactic_2003} initial mass function.

\section{Simulated submillimetre galaxies}
\label{sec:sims}

\subsection{Simba}
\textsc{Simba}\ \citep{dave_simba:_2019} is the successor to the \textsc{Mufasa} simulations \citep{dave_mufasa:_2016,dave_mufasa:_2017}.
Both simulations use the Meshless Finite Mass (MFM) method from \textsc{Gizmo} \citep{hopkins_new_2015}, but \textsc{Simba}\ implemented a number of significant improvements to the sub-grid prescriptions for both star formation and AGN feedback. \textsc{Simba}\ was tuned primarily to match the evolution of the overall stellar mass function and the stellar mass--black hole mass relation \citep{dave_simba:_2019}, but reproduces a number of key observables at both low and high redshift that do not rely on this tuning, and are bona fide predictions of the model.
These include SFR functions, the cosmic SFR density, passive galaxy number densities \citep{rodriguez_montero_mergers_2019}, galaxy sizes and star formation rate profiles \citep{appleby_impact_2020}, central supermassive black hole properties \citep{thomas_black_2019}, damped Ly$\alpha$ abundances \citep{hassan_testing_2020}, star formation histories \citep{mamon_frequency_2020}, the reionisation-epoch UV luminosity function \citep{wu_photometric_2020}, and the low-redshift Ly$\alpha$ absorption \citep{christiansen_jet_2019}. We refer the reader to \citet{dave_simba:_2019} for a full description of the simulation, and to the works referenced above for details of the specific predictions and comparison to observational constraints. Of importance to this work is the treatment of dust in \textsc{Simba}, which we briefly summarise here.

\textsc{Simba}\ implements a self-consistent on-the-fly dust framework, modelling the production, growth and destruction of dust grains, passively advected along gas elements \citep{dave_simba:_2019,li_dust--gas_2019}.
Metals ejected from SNe and AGB stars condense into grains following the \cite{dwek_evolution_1998} prescription, and these grains are assumed to have a single size, 0.1\mumetre. Condensation efficiencies for each process are updated based on the theoretical models of \cite{ferrarotti_composition_2006} and \cite{bianchi_dust_2007}. Two-body processes can increase the amount of dust by accreting gas-phase metals \citep{dwek_evolution_1998,hirashita_dust_2000,asano_dust_2013}, and `thermal sputtering' and SNe shocks \citep{mckinnon_dust_2016} can destroy grains.
A number of processes can completely destroy the dust reservoir in a gas element; these include hot-phase winds, star formation and any gas subject to X-ray or jet feedback from AGN. This prescription results in dust-to-metal ratios in good agreement with observations, and dust mass functions broadly in agreement with data, albeit somewhat low at $z\sim 2$~\citep{li_dust--gas_2019}; \textsc{Simba}\ may mildly underestimate the dust content of dusty SFGs during Cosmic Noon, an epoch of interest in this work. Nevertheless, \cite{lovell_reproducing_2021} show that \textsc{Simba} can broadly reproduce the demographics of the SMG population, again with no tuning, with the best match to observational number counts compared to any other fully hydrodynamic cosmological simulation to date.

In this work we primarily use a large volume $(147 \,\mathrm{cMpc})^{3}$ \textsc{Simba} run with the fiducial physics in order to study rare, massive SMGs. This volume contains 1024$^3$ dark matter particles and 1024$^3$ gas elements, with element (particle) masses of $6.3\times 10^7 M_\odot$ and $1.2\times 10^7 M_\odot$, respectively, and an adaptive gravitational softening length covering 64 neighbours with a minimum value of $0.5 \,h^{-1}\,\mathrm{kpc}$.
Galaxies are identified using an on-the-fly friends-of-friends structure finder, described in \cite{dave_mufasa:_2016}.
This means that individual SMGs are resolved with thousands of gas elements at minimum.

\subsection{Dust continuum radiative transfer}

\begin{figure*}
	\includegraphics[width=\textwidth]{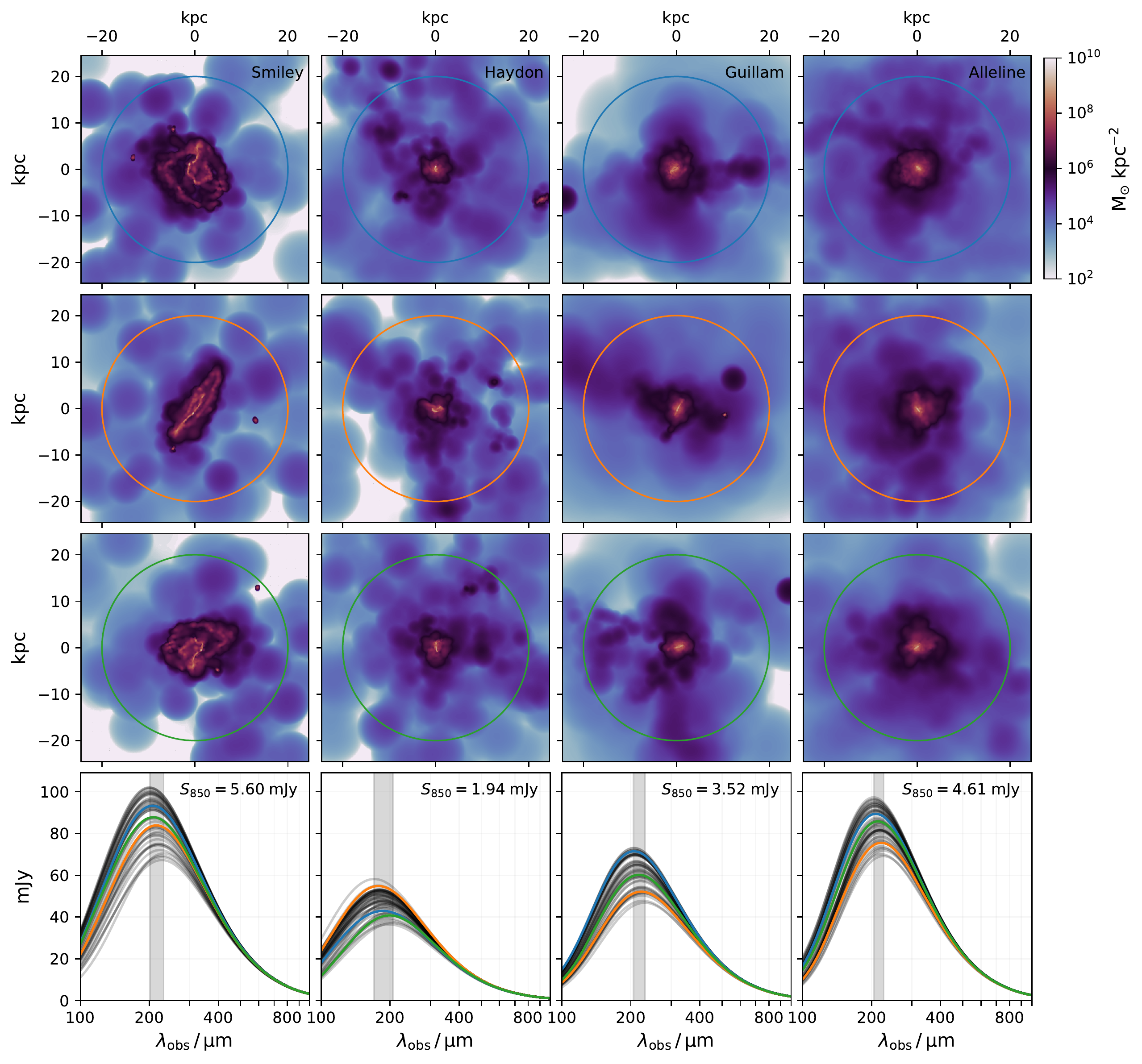}
    \caption{\textit{Top three rows:} dust column density maps for each galaxy.
    Each column shows an individual galaxy (Smiley, Haydon, Guillam, Alleline, labelled top right), and each row shows one of three orthogonal orientations, colour coded by a circle with radius 20 kpc.
    \textit{Bottom row:} observer-frame ($z = 2.025$) SED in the sub-mm regime for all 50 random orientations in black.
    The SEDs for the 3 orientations mapped above are also shown; the line colour corresponds to the colour of the circles.
    The median flux density at 850 \mumetre (observer-frame) is given at the top right.
		The vertical shaded region shows the range in wavelength of the location of the peak of the sub-mm emission.
    }
    \label{fig:dust_maps}
\end{figure*}

\begin{figure*}
	\includegraphics[width=\textwidth]{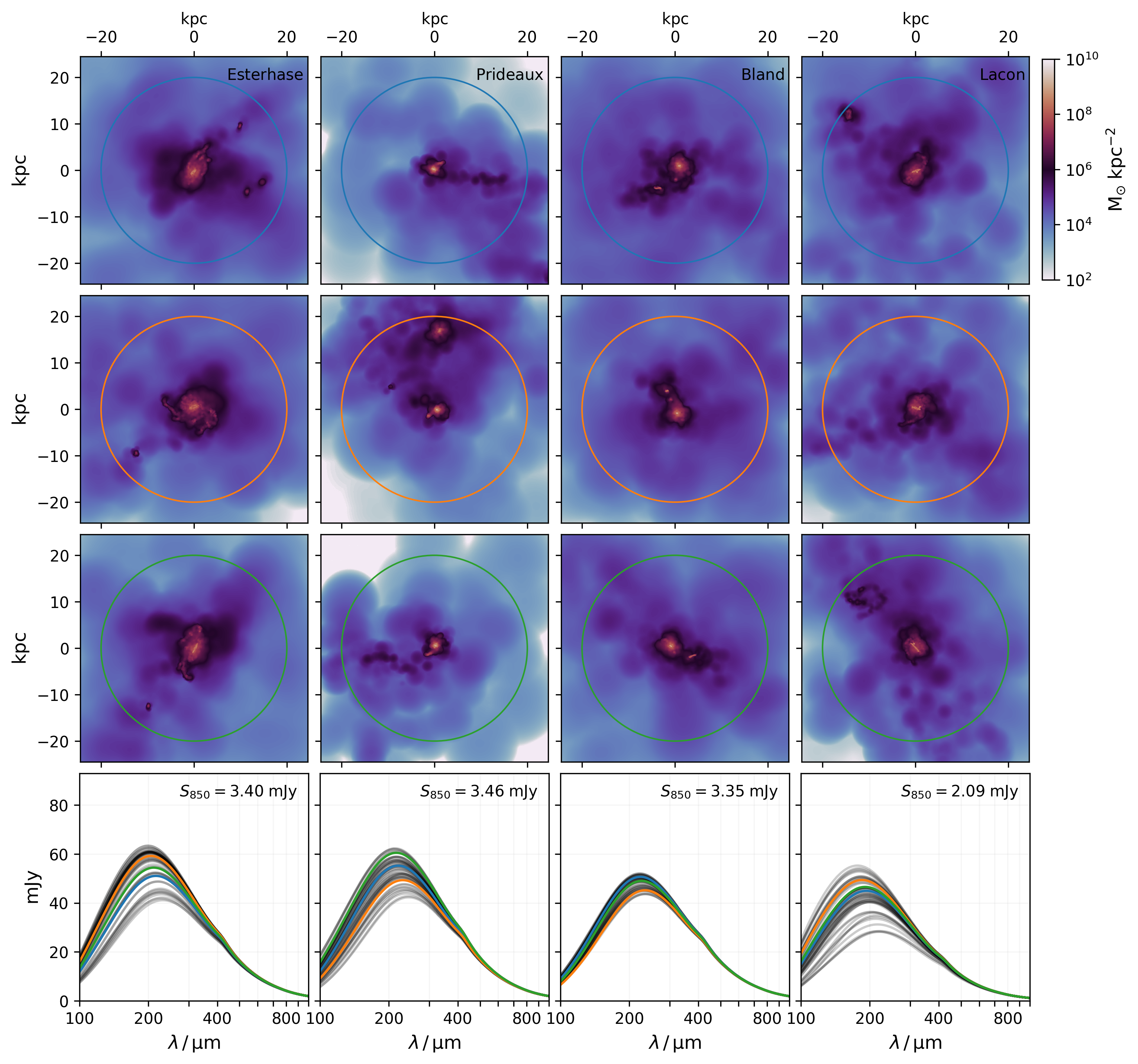}
    \caption{As for \fig{dust_maps}, but showing Esterhase, Prideaux, Bland and Lacon.}
    \label{fig:dust_maps_B}
\end{figure*}

We use the {\it Powderday} dust radiative transfer (RT) code \citep{narayanan_powderday_2021}. We refer the reader to \cite{narayanan_powderday_2021} and \cite{lovell_reproducing_2021} for a comprehensive description of the RT methodology.
In summary, we generate full SEDs for galaxies as follows.
Star particles are treated as a Simple Stellar Population (SSP) with fixed age and metallicity.
The Flexible Stellar Population Synthesis model \citep[FSPS;][]{conroy_propagation_2009,conroy_propagation_2010} is used to generate an SED, assuming an initial mass function and theoretical isochrone library.
We adopt the MILES spectral library \citep{sanchez-blazquez_medium-resolution_2006} combined with BPASS  \citep{eldridge_binary_2017,stanway_re-evaluating_2018} to take into account binary evolution.
Radiation from each source propagates through the ISM where photons are scattered, absorbed and re-emitted.
We assume the wavelength-dependence of dust RT is described by the \cite{draine_interstellar_2003} model, with $R_V=3.1$.
Subgrid RT processes below the simulation resolution, for example the effect of dust in the photodissociation region surrounding star forming cores, are not modelled, as this would introduce a significant number of extra parameters into our model.
We also argue that such processes would have a small impact on the macro-level orientation effects explored in this work.
{\it Hyperion} \citep{robitaille_hyperion_2011} is used to perform dust RT using a Monte Carlo approach.
Emission from each source is modelled with photon packets, which are emitted with random direction and frequency.
Photons propagate through the volume until they reach the edge of the grid, or some limiting optical depth $\tau_{\rm lim}$.
Finally, emergent SEDs are generated through ray tracing.
We restrict our analysis to a cube with diameter 30 kpc (physical) centred on each galaxy.

\subsection{Sample}

In order to focus on the effect of orientation we select a single snapshot, at redshift $z = 2.025$, similar to the median redshift of 850 \mumetre-selected SMGs \citep{chapman_redshift_2005,simpson_alma_2014,dudzeviciute_alma_2020}.
We select all galaxies with more than 1000 star particles and a total star formation rate within the halo of $\mathrm{SFR} > 500 \, \mathrm{M_{\odot} \, yr^{-1}}$
This results in eight galaxies\footnote{Named in homage to John Le Carr\'e, who sadly passed away during the preparation of this paper.}, listed in \tab{sample}.\footnote{SFRs are quoted within a 30 kpc aperture centred on the galaxies centre of mass, hence why some of these galaxies have SFRs lower than our halo-defined selection criteria.}
These galaxies are sub-mm luminous ($S_{850}\approx2\text{--}6$\,mJy), occupying the bright end of the $S_{850}$ number count distribution in \simba\ at this redshift \citep{lovell_reproducing_2021}.
Figures \ref{fig:dust_maps} and \ref{fig:dust_maps_B} show the dust column density distribution in the selected galaxies along three orthogonal sightlines.

\begin{table*}
    \centering
    \begin{tabular}{lcccccccc}
    \hline
    Label & $\mathrm{log_{10}}(M_{\star,30 \; \mathrm{kpc}} \,/\, \mathrm{M_{\odot}})$ & $\mathrm{log_{10}}(M_{\mathrm{dust}} \,/\, \mathrm{M_{\odot}})$ & $\mathrm{SFR \,/\, M_{\odot} \, yr^{-1}}$ & $\left< S_{850} \right> \,/\, \mathrm{mJy}$ & $T_{\mathrm{MBB}} \,/\, \mathrm{K}$ & $T_{\mathrm{peak}} \,/\, \mathrm{K}$ & $\mathrm{log_{10}}(L_{\mathrm{IR}} \,/\, \mathrm{L_{\odot}})$ & Disc? \\
    \hline
    Smiley & 12.03 & 9.29 & 1238 & 5.60 & $49.25_{-3.39}^{+1.40}$ & $53.69_{-2.52}^{+1.22}$ & $13.03_{-0.10}^{+0.04}$ & Yes \\[2pt]
    Haydon & 11.83 & 8.90 & 618 & 1.94 & $57.97_{-3.08}^{+2.13}$ & $60.11_{-3.98}^{+1.74}$ & $12.76_{-0.05}^{+0.04}$ & No \\[2pt]
    Guillam & 11.46 & 8.98 & 432 & 3.52 & $50.92_{-3.49}^{+1.96}$ & $48.91_{-1.07}^{+1.38}$ & $12.89_{-0.10}^{+0.04}$ & Yes \\[2pt]
    Alleline & 11.45 & 9.08 & 962 & 4.61 & $51.30_{-1.78}^{+2.33}$ & $48.83_{-0.96}^{+1.57}$ & $13.02_{-0.05}^{+0.04}$ & No \\[2pt]
    Esterhase & 11.34 & 8.95 & 719 & 3.41 & $53.02_{-4.66}^{+0.84}$ & $52.01_{-3.77}^{+0.74}$ & $12.90_{-0.10}^{+0.02}$ & Yes \\[2pt]
    Prideaux & 11.33 & 8.85 & 736 & 3.30 & $49.94_{-2.56}^{+1.56}$ & $48.42_{-2.05}^{+1.67}$ & $12.86_{-0.06}^{+0.04}$ & No \\[2pt]
    Bland & 11.26 & 8.87 & 641 & 3.27 & $49.14_{-1.37}^{+0.79}$ & $47.14_{-1.18}^{+0.83}$ & $12.82_{-0.03}^{+0.02}$ & No \\[2pt]
    Lacon & 11.26 & 8.75 & 467 & 2.08 & $57.63_{-4.00}^{+3.06}$ & $54.69_{-2.44}^{+2.77}$ & $12.78_{-0.08}^{+0.05}$ & Yes \\[2pt]
    \hline
    \end{tabular}
 \caption{The properties of the galaxy sample. The columns, from left to right, show (1) stellar mass, (2) total dust mass, (3) instantaneous star formation rate, (4) median 850 \mumetre\ (observer-frame) flux density (over all orientations), (5,7) median and $5^{\mathrm{th}} - 95^{\mathrm{th}}$ percentile of the temperature and FIR luminosity from a modified blackbody fit to the sub-mm emission over the 50 orientations (see \sec{uncertain}), (6) median and $5^{\mathrm{th}} - 95^{\mathrm{th}}$ percentile of the temperature distribution measured from the peak of the FIR emission, (7) whether the galaxy exhibits a disc-like morphology, from visual inspection of the 3D dust and SFR distribution.
 (1,2,3) are all measured within a 30 kpc spherical aperture centred on the halo centre of mass.}
 \label{tab:sample}
\end{table*}

\section{Analysis}
\label{sec:analysis}

\begin{figure}
	\includegraphics[width=\columnwidth]{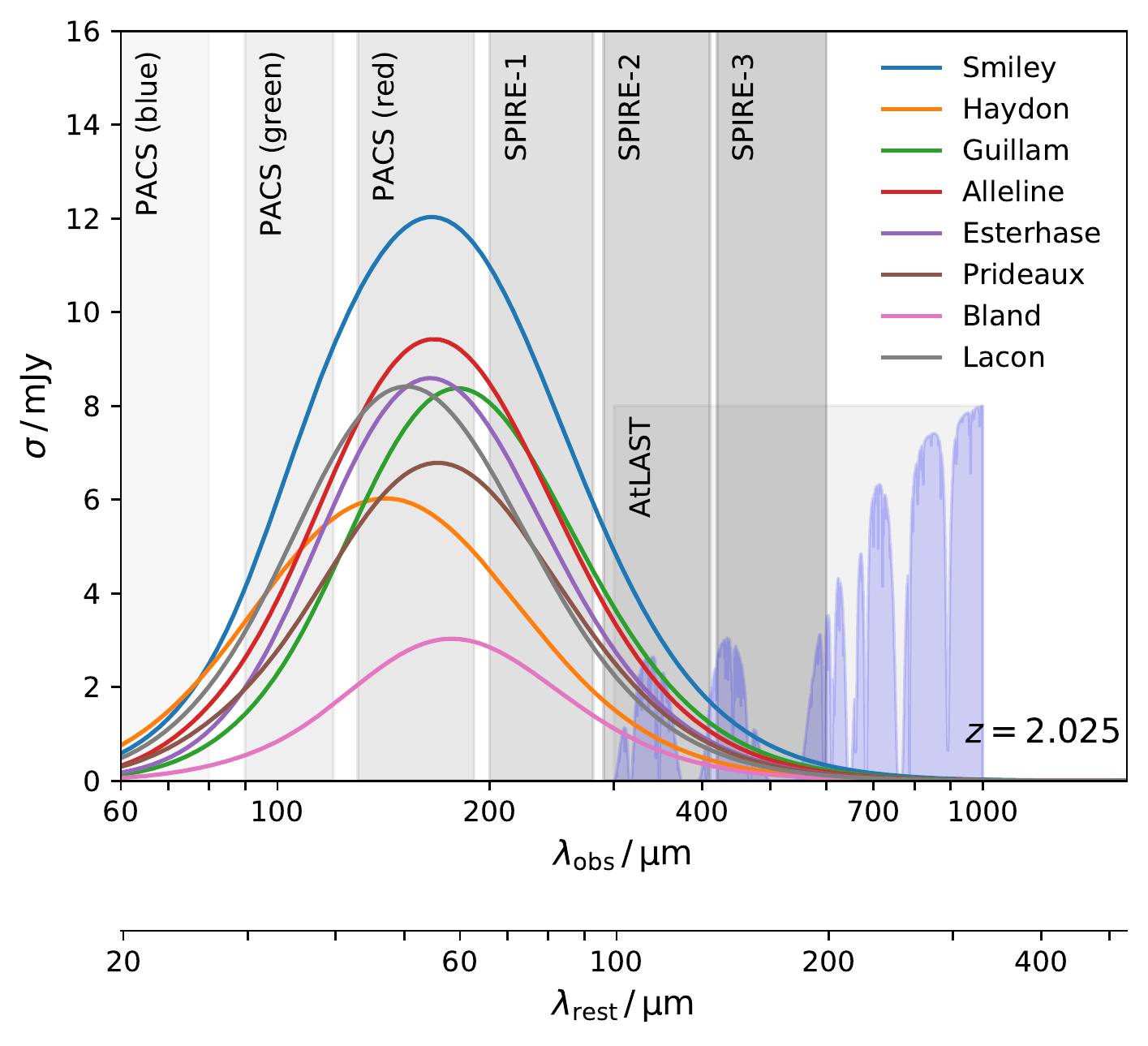}
    \caption{
        Standard deviation of the flux density over all 50 orientations, for each galaxy, as a function of (observer-frame) wavelength (the equivalent rest-frame wavelength is shown for reference).
				The wavelength ranges covered by the Herschel PACS and SPIRE instruments are shown for reference \citep{poglitsch_photodetector_2010,griffin_herschel-spire_2010}, as well as the anticipated wavelength range of the AtLAST instrument \citep{kawabe_new_2016,groppi_first_2019,klaassen_atacama_2020}. The atmospheric transmission in the AtLAST wavelength regime is shown in blue (taken from \href{https://almascience.eso.org/about-alma/atmosphere-model}{almascience.eso.org/about-alma/atmosphere-model})
    }
    \label{fig:sigma_sed}
\end{figure}

For each galaxy we generate the galaxy-integrated SEDs as observed by 50 coeval observers distributed randomly throughout the $z=0$ Universe.
Figures \ref{fig:dust_maps} \& \ref{fig:dust_maps_B} show the variety of \textit{observed} 100--1000 \mumetre\ SEDs for each galaxy in the sample.
This illustrates substantial variance in the observed SED close to the peak of the thermal dust emission simply due to projected orientation.
This variation at the peak is considerable, a factor of 2 for Smiley, and up to a factor of 2.7 for Lacon.
On average, the variation is of factor 1.9 for all our simulated galaxies.
\fig{sigma_sed} shows the standard deviation, as a function of wavelength, for the flux density across all 50 orientations.
Each galaxy shows a similar increase in the variability towards the peak of the dust emission, though with different levels of absolute variation ($>$10\,mJy for Smiley at $\lambda_{\rm obs}\sim$180 \mumetre).
The SEDs converge towards millimetre wavelengths as expected for optically thin radiation in the Rayleigh-Jeans regime.

There are four (linked) implications of this result:
\begin{enumerate}
    \item A single random observer will never directly measure the representative SED of a given SMG.

    \item As a result of (i), at a fixed redshift, derived properties such as dust temperature, luminosity and SFR are subject to an additional source of uncertainty which we will refer to as the `orientation uncertainty'.

    \item For a flux-limited survey conducted at $\lambda\lesssim200$ \mumetre\ (rest-frame), the selection of SMGs at a fixed redshift will be biased in favour of certain projected orientations, which we will refer to as the `orientation bias'.

		\item Such a flux-limited survey will also be subject to an additional redshift bias, as the \textit{intrinsic} wavelength window in which orientation effects are most prominent moves through the \textit{observed} wavelength range.

\end{enumerate}
Below, we explore these four points in more detail.
However, we first investigate the origin of the orientation dependence.

\subsection{Explaining the orientation dependence}
\label{sec:explain}

The only property that changes in our SMGs when observed from different orientations is the projected geometry of both the stellar sources and dust distribution.
The RT pipeline self consistently takes this complex geometry into account.
We now investigate whether this orientation-dependent variation can be simply parameterised by the physical properties of the galaxies.

It is clear from Figures \ref{fig:dust_maps} \& \ref{fig:dust_maps_B} that a considerable fraction of our SMGs have disc like morphologies.
To explore whether this may lead to the orientation-dependent variation, we first calculated the angular momentum vector of each of our galaxies gas distributions,
\begin{align}
    \boldsymbol{L} = \sum_i \left( \boldsymbol{R}_i \times m_i \boldsymbol{V}_i \right) \;\;,
\end{align}
where $\boldsymbol{R}_i$ is the position of gas particle $i$, $m$ is its mass, and $\boldsymbol{V}_i$ its velocity.
We restricted this calculation to particles within 100\,kpc (physical) of the centre of mass.
The angular momentum vector points outwards from the position of the centre of mass, perpendicular to the plane of the disc (if present).
To find the orientation of the galaxy with respect to our chosen line of sight, we then calculated the cosine similarity, $C$, between the line of sight vector, $\boldsymbol{u}$, and $L$,
\begin{align}
    C = \frac{\boldsymbol{L} \cdot \boldsymbol{u}}{\lVert L \rVert \, \lVert u \rVert}
\end{align}
where $C \in [-1,1]$. $|C| = 1$ suggests alignment of the two vectors, i.e., a face-on view of a galaxy, and $|C| = 0$ suggests an edge-on view.

\begin{figure}
	\includegraphics[width=\columnwidth]{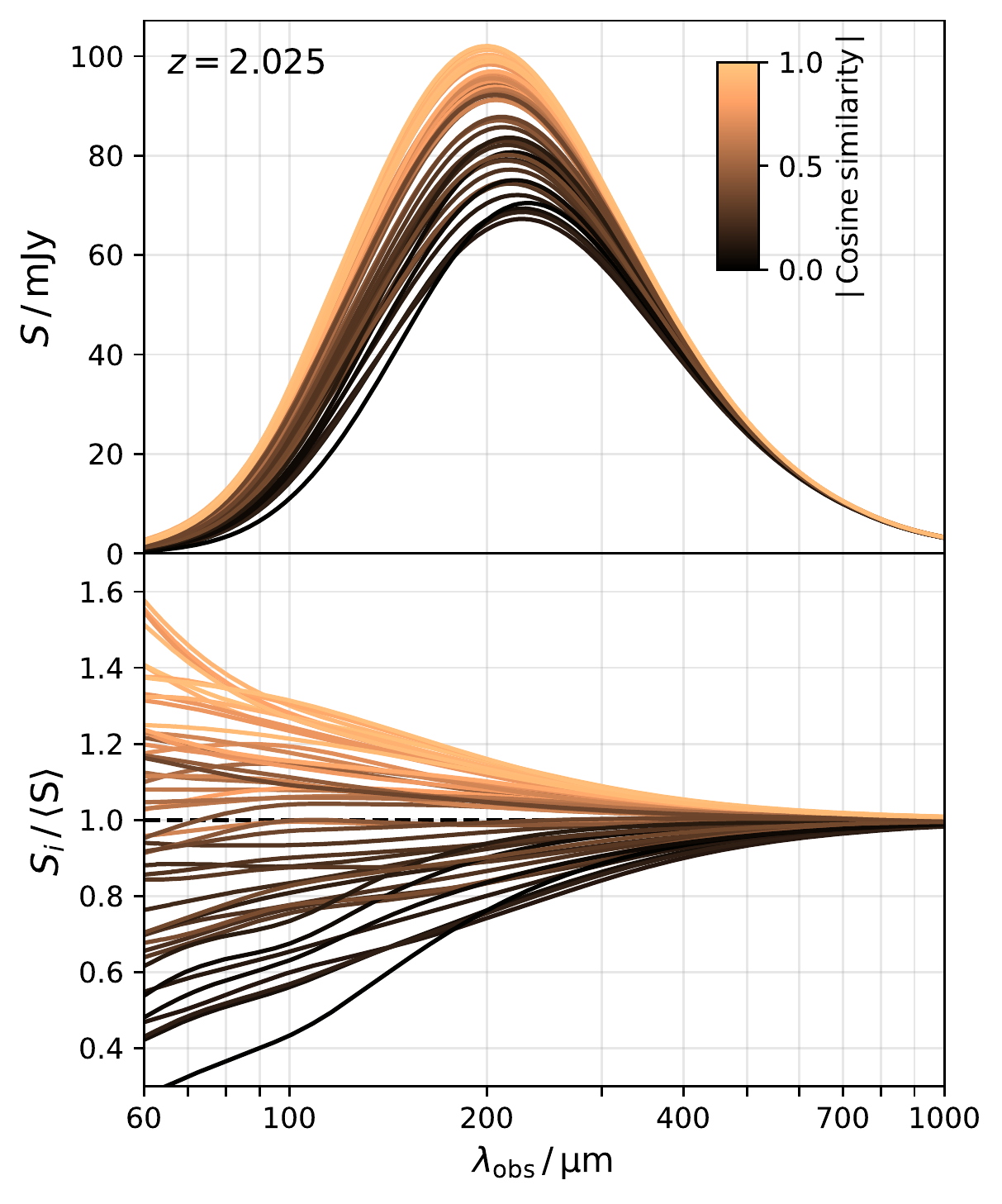}
    \caption{\textit{Top:} observer-frame spectral energy distribution for Smiley, coloured by the absolute value of the cosine similarity ($|C|$) between the angular momentum vector of the gas and the orientation vector.
    \textit{Bottom:} ratio of the flux density for each orientation to the mean over all orientations.
    }
    \label{fig:cosine_similarity_g3}
\end{figure}

\begin{figure}
	\includegraphics[width=\columnwidth]{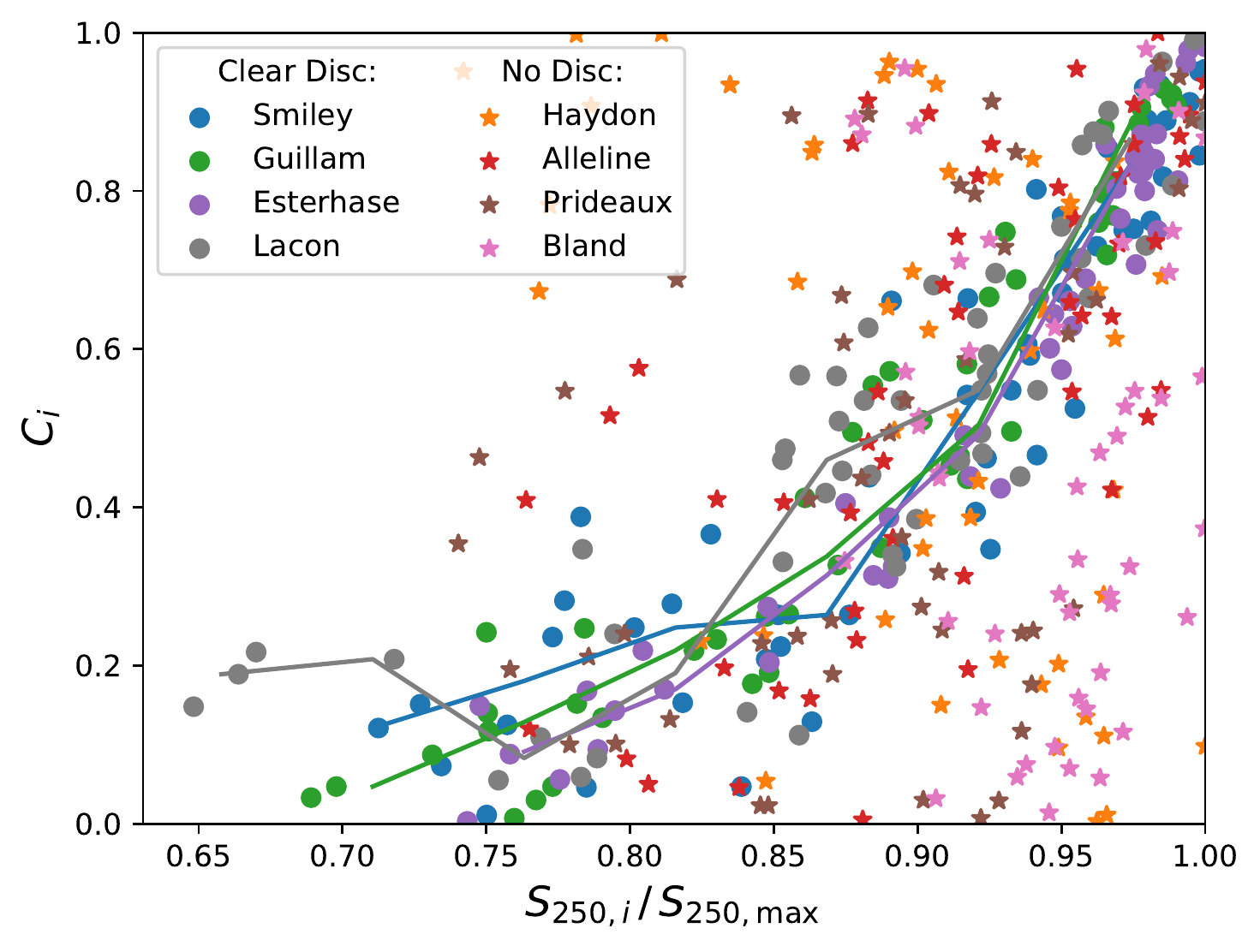}
    \caption{Cosine similarity $C$ against $S_{250}$ flux density (observer-frame), normalised to the maximum of the 50 orientations.
    Galaxies with clear disc morphologies (Smiley, Guillam, Esterhase, Lacon; circular markers) show a clear positive correlation, further emphasised by the lines which show the binned medians, whereas galaxies without a disc (Haydon, Alleline, Prideaux, Bland; star markers) show no such relationship.
    }
    \label{fig:cosine_similarity}
\end{figure}

\fig{cosine_similarity_g3} shows the sub-mm emission in Smiley for all 50 orientations, coloured by the cosine similarity.
There is a clear dependence of the normalisation in the optically-thick regime on $C$: where the disc is viewed edge-on, the emission is lowest, and where the disc is viewed face-on the the emission is highest.
In order to compare all galaxies in our sample we show $C$ against the normalised flux density in \fig{cosine_similarity}.
For galaxies with clear disc-like morphologies (Smiley, Guillem, Esterhase, Lacon) there is a clear correlation between $C$ and the normalised flux density.
However, not all sub-mm galaxies in our sample have disc morphologies (Haydon, Alleline, Prideaux, Bland); in these cases the angular momentum vector is meaningless, and there is no correlation.

We wish to find a morphology-agnostic measure relating the normalised flux-density to the orientation.
For FIR emission in the optically-thick regime ($\tau > 1$), the emission is from the surface of the dust.
It may be expected that the flux density will therefore be proportional to the projected area of this $\tau > 1$ surface, for some given orientation.
It is computationally expensive to generate maps of the emission for all 50 orientations of each of our selected galaxies; another way of estimating the size of this surface is to use the dust column density to infer maps of the optical depth.
We computed this for all 50 orientations of our selected galaxies and found that, as for the cosine similarity, only disc galaxies showed a strong correlation between the size of the $\tau > 1$ surface and the flux density.

\begin{figure*}
	\includegraphics[width=\textwidth]{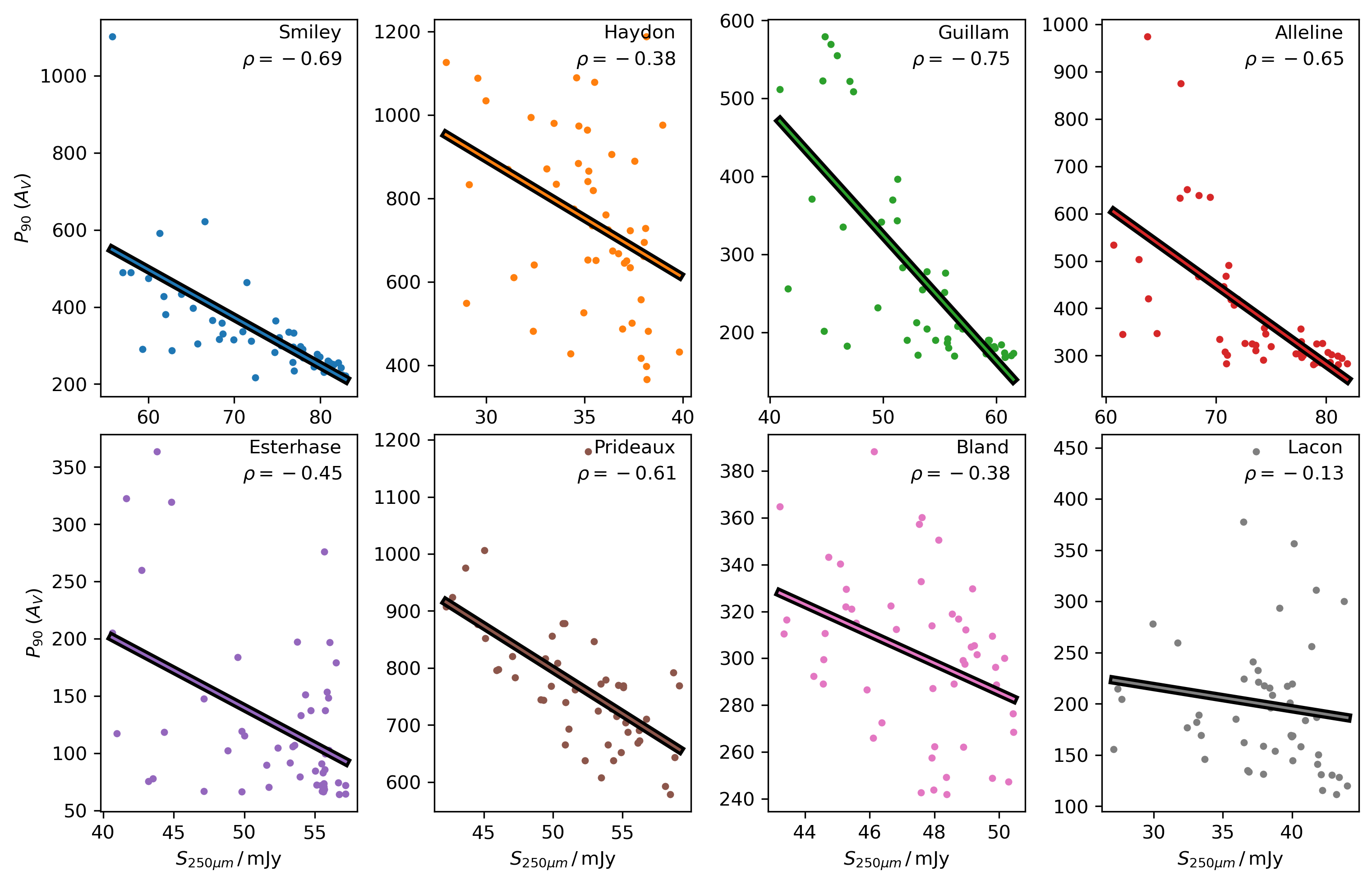}
    \caption{
    The $90^{\mathrm{th}}$ percentile of the extinction distribution (to each gas particle in the galaxy) against the 250$\mathrm{\mu m}$ flux density, for each of 50 random orientations of each galaxy. Each panel shows a different galaxy from the sample, with a linear relation plotted, and the pearson correlation coefficient $\rho$ quoted.
    }
    \label{fig:A_V_gas_p90}
\end{figure*}

A more realistic means of estimating optical depth maps is to compute the line-of-sight (LOS) column density of dust, and relate this to the extinction.
This is typically calculated towards individual stellar particles in a galaxy, by summing the contributions from each gas particle whose SPH kernel overlaps a line between the observer and the chosen particle, in order to approximate the effect of dust attenuation on the emission.
However, given that we wish to understand the effect on the \textit{dust} emission, we instead calculate the LOS column density towards each \textit{gas} particle.
We then explored the correlations between the derived extinction and the normalised flux density.
\fig{A_V_gas_p90} shows the $\mathrm{90^{th}}$ percentile of the extinction distribution for each galaxy orientation against the $250 \mathrm{\mu m}$ emission.
All of our galaxies show a negative correlation between the extinction and the emission, for a given orientation, and this is strongest for disc galaxies.
These results highlight the impact of the complex geometry of the gas and stars on the emergent emission, emphasising the need for full RT to explore this effect in simulations.




%

\subsection{Orientation uncertainty}
\label{sec:uncertain}

We have demonstrated the complex impact of orientation on the observed FIR emission of a galaxy.
We now explore how this affects measured derived properties.
For galaxies detected in the sub-mm, the most important derived properties are the temperature and integrated infrared luminosity, typically estimated through modified blackbody (MBB), or `greybody', fits to photometric sampling of the sub-mm SED. Temperature is a parameter of the model, and the luminosity is simply measured through the integral of the MBB.
Further derived properties such as dust/ISM mass and SFR can follow \citep[e.g.,][]{kaasinen_molecular_2019}. 
Here we illustrate the impact of orientation uncertainty on estimates of dust temperature and luminosity from MBB fits, noting that this uncertainty will propagate further into the other derived measurements such as dust/ISM mass and SFR. The use of a modified blackbody in fitting far-infrared SEDs is {\it de rigueur}, and considerations of the impact of optical depth effects in the analysis of SMGs is certainly not new.
For example, \cite{simpson_scuba-2_2017} used ALMA to measure the sizes of SMGs selected from the SCUBA-2 Cosmology Legacy Survey \citep{geach_scuba-2_2017}, arguing that their far-infrared emission arises from regions that are optically thick at $\lambda_0\gtrsim 75$\,$\mu$m, and discuss the impact this has on the measured dust temperature \citep[see also][]{dudzeviciute_alma_2020}.

\begin{figure}
	\includegraphics[width=\columnwidth]{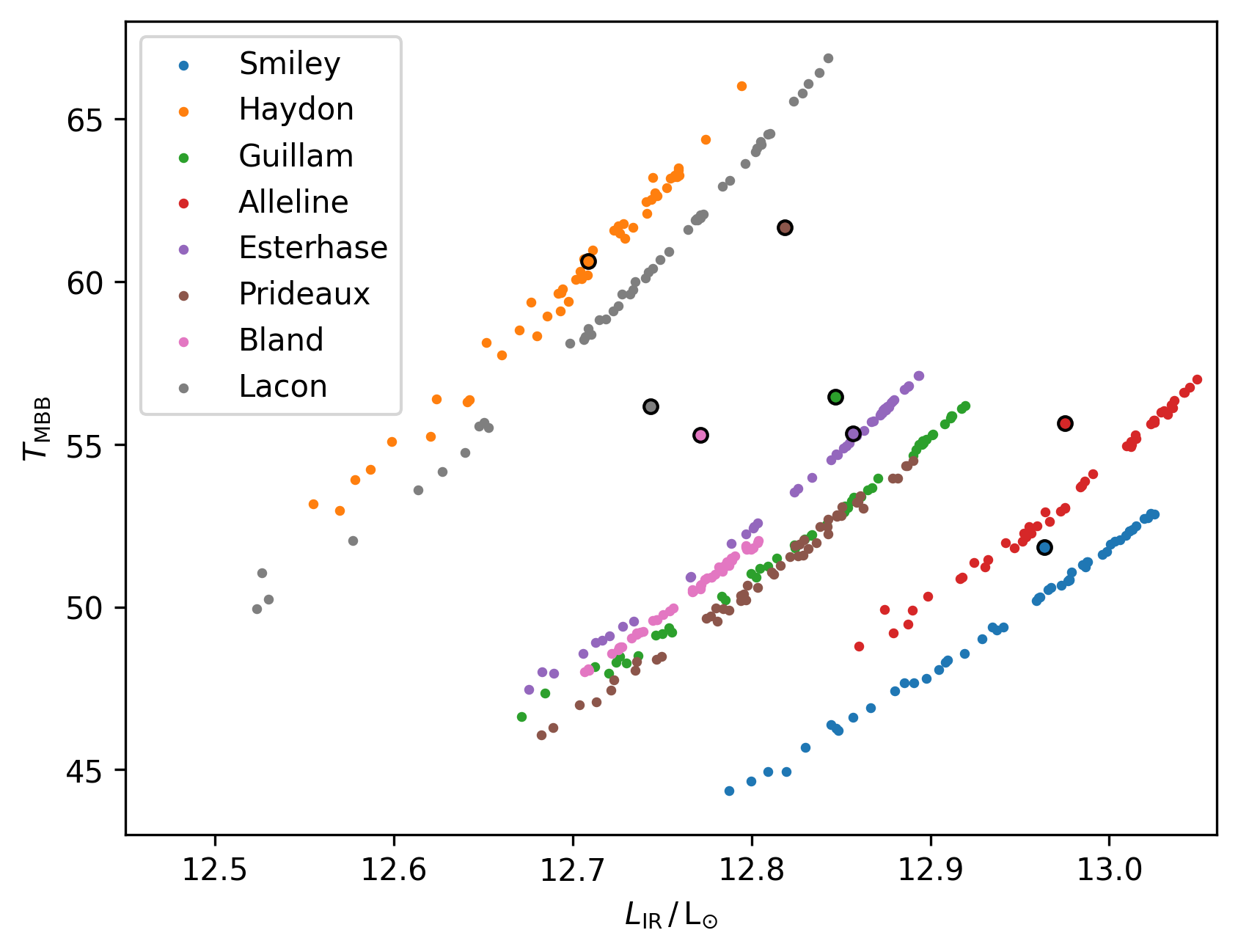}
    \caption{
		Distribution of estimates of temperature and FIR luminosity for all 50 orientations of our eight sub-mm galaxies.
		\textit{Top:} Temperature ($T_{\mathrm{MBB}}$) and $L_{\mathrm{IR}}$ estimated through MBB fits to the sub-mm emission.
		\textit{Bottom:} The temperature estimate from the peak of the SED in the FIR regime ($T_{\mathrm{peak}}$).
    }
    \label{fig:temp_lum_scatter}
\end{figure}

We generally do not fully sample the SED in practice.
With photometry in the sub-mm we may gather a handful of measurements of the total flux density, ideally spanning the peak of the thermal dust emission and into the Rayleigh-Jeans tail.
The sub-sampled SED can then be modelled with a MBB \citep{casey_far-infrared_2012}, $B_\nu(T)$, of the form
\begin{align}
    B_\nu  = \frac{\left(1 - e^{\tau(\nu)}\right)\, \nu^{3}}{e^{h \nu / k T_{\mathrm{MBB}}} - 1} \,\,,
\end{align}
where $h$ is the Planck constant, $k$ the Boltzmann constant, $\tau$ is the optical depth given by $\tau = (\nu / \nu_{0})^{\beta}$, and $\beta$ is the dust emissivity.
The transition between the optically thick and thin regimes occurs at $\tau(\nu_0)\approx1$.
Given the emission at $[250,350,500,850]$\mumetre\ (which a sub-mm observer would consider a well-sampled SED) we then fit for the temperature $T_{\mathrm{MBB}}$, with some arbitrary normalisation factor.
As discussed in \cite{liang_dust_2019}, this should not be considered a physical dust temperature.
Note that we assume the redshift of the galaxy is known, allowing us to investigate the impact of orientation uncertainty independent of redshift uncertainty.
In practice, the redshifts of SMGs can be routinely measured with high precision through sub-mm/mm emission line spectroscopy \citep[{e.g.,}][]{strandet_redshift_2016}.
We also fit $\nu_0$ and $\beta$ simultaneously with $T_{\mathrm{MBB}}$.
We find $\lambda_{0} = c / \nu_0 = 80 \, \mu \mathrm{m}$, low compared to the standard assumption \citep[\textit{e.g.}][]{blain_submillimetre_2003} but consistent with recent ALMA measurements at high-$z$ \citep{faisst_alma_2020}.
We also obtain $\beta = 2$, consistent with some observational constraints \citep[\textit{e.g.}][]{dunne_scuba_2000,draine_dust_2007,magnelli_herschel_2012}.

The top panel of \fig{temp_lum_scatter} shows the combined distribution of $T_{\mathrm{MBB}}$ and $\log_{10}(L_{\rm FIR}/L_\odot)$ (the latter obtained from integrating the total MBB SED over 1--1000\mumetre) for each galaxy over the 50 orientations.
\tab{sample} shows the median and $5^{\mathrm{th}} - 95^{\mathrm{th}}$ percentile range of the distributions for each galaxy.
The measured temperatures are somewhat high compared to observed galaxies at $z = 2$ \citep{schreiber_dust_2018}, though within the considerable scatter at fixed $L_{\mathrm{IR}}$, particularly for the most IR luminous sources sampled here, and are consistent with those obtained from the previous MUFASA model \citep{narayanan_irx-_2018}.
The average inter-percentile range is 5.0 K and $0.110  \; \mathrm{dex}$, for the temperature and FIR luminosity, respectively.
For a simple SFR calibration \citep[\textit{e.g.}][]{kennicutt_star_2012} based on $L_{\rm FIR}$, this corresponds to a 30\% variance in SFR, comparable to the systematic uncertainty due to choice of initial mass function.
Our results quantify the added uncertainty optical depth effects have due to the relative orientation of the source.
As fixed observers we only have a single view of any given external galaxy; the orientation-dependent variance on $T$ and $L_{\rm FIR}$ represent fundamental uncertainties to be considered when interpreting observations of SMGs, even those with exquisite data.

Whilst MBB fits are used frequently in the fitting, analysis and interpretation of far-IR SEDs, it has been demonstrated that there are a number of significant degeneracies between the fitted parameters (\textit{e.g.} anti-correlation between $T$ and $\beta$) when applied to real data \citep[see][]{sajina_1-1000m_2006,shetty_effect_2009,shetty_effect_2009-1,papadopoulos_co_2010}.
To avoid the added uncertainties introduced by the MBB modelling, we also show the temperature measured from the peak of the far-infrared emission \citep[\textit{e.g.}][]{casey_dusty_2014,liang_dust_2019},
\begin{equation}
	T_{\mathrm{peak}} = \frac{2.90 \times 10^3 \, \mathrm{\mu m \cdot K} }{ \lambda_{\mathrm{peak}}}
\end{equation}
where we infer $\lambda_{\mathrm{peak}}$ directly from the fully sampled output SED, and use quadratic interpolation to find the exact position of the peak.
We also measure $L_{\rm FIR}$ directly from the output SED.
The bottom panel of \fig{temp_lum_scatter} shows the scatter in these directly calculated parameters over the 50 orientations.
$T_{\mathrm{peak}}$ and $T_{\mathrm{MBB}}$ are broadly similar, though there is a weaker correlation between $T_{\mathrm{peak}}$ and $L_{\rm FIR}$ than for the MBB inferred temperatures, though the quantitative scatter in each parameter is similar.
\tab{sample} shows the median and $5^{\mathrm{th}} - 95^{\mathrm{th}}$ percentile range of the distribution of $T_{\mathrm{peak}}$ for each galaxy.


\subsection{Orientation dependent survey bias}
\label{sec:bias}

We have shown how orientation can lead to a large dispersion in the emergent rest-frame submillimetre flux density, and its effect on measured intrinsic properties.
We introduce the concept of the `orientation factor', which describes the combined effect of orientation and morphology on the relative emission.
We now explore how this can lead to a bias in the selection of sub-mm galaxies in flux-limited surveys, by preferentially selecting galaxies with certain orientations, and the knock--on effect on measured distributions of intrinsic quantities.

To demonstrate the magnitude of such a bias we build a Monte Carlo model for a sub-mm survey at a given observed wavelength $\lambda_{\mathrm{obs}}$.
There are three main components to the model: the distribution of the `orientation factor' as a function of rest--frame wavelength ($\lambda_{\mathrm{rest}}$), a number density distribution of sources selected at $\lambda_{\mathrm{obs}}$, and a redshift distribution of those sources selected at $\lambda_{\mathrm{obs}}$.
The latter component takes account of the fact that galaxies observed at a single observer--frame wavelength cover a range of rest--frame wavelengths depending on their distance.
Whilst the negative $K$--correction leads to consistent flux densities over a large range of redshift, the variability due to orientation at a given redshift, at a fixed observer--frame wavelength, will change, and lead to selection biases as a function of redshift.
The model is publicly available at \href{https://github.com/christopherlovell/orientation_bias}{github.com/christopherlovell/orientation\_bias}, and can be used to assess the impact of orientation completeness for an arbitrary survey.

\begin{figure}
	\includegraphics[width=\columnwidth]{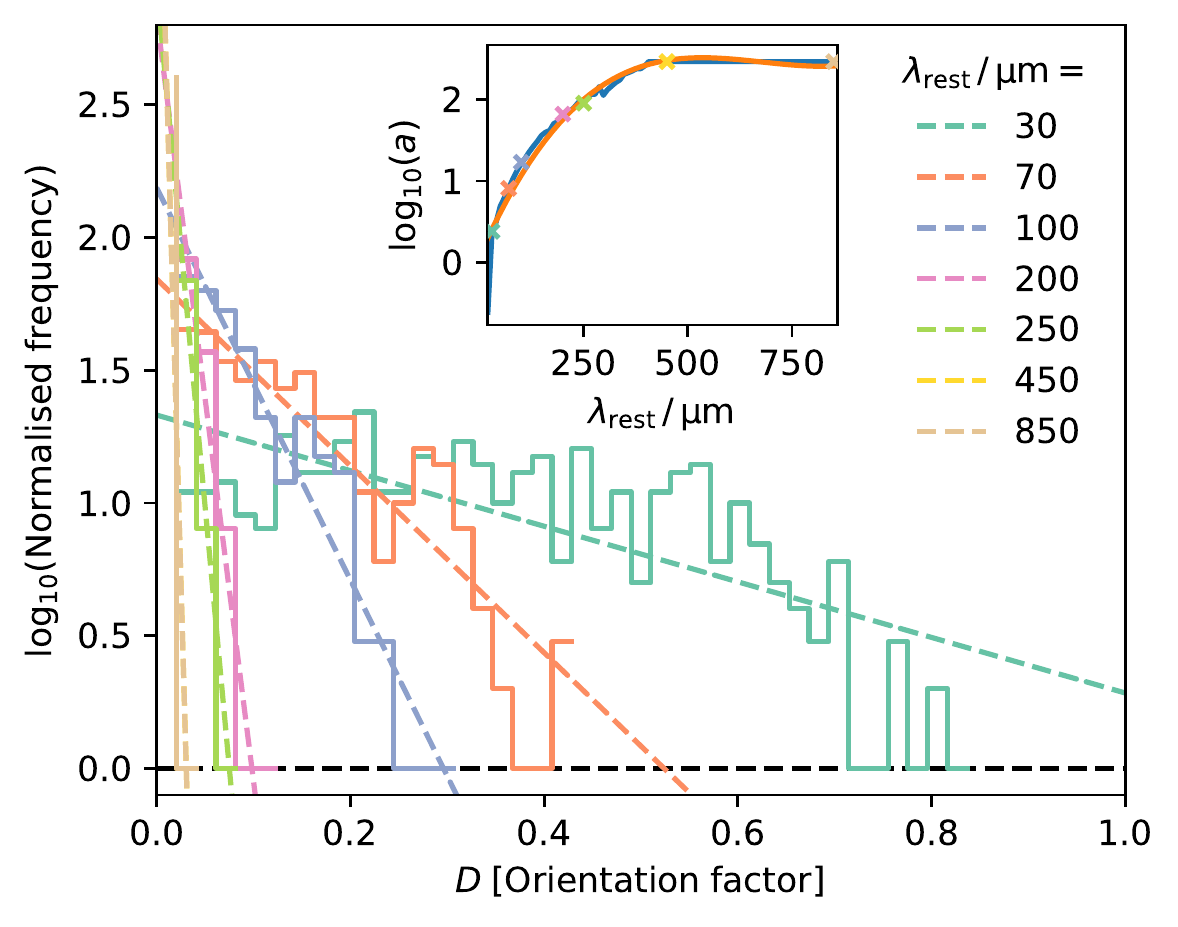}
    \caption{
    Binned distribution of orientation factor for all galaxies and orientations combined (solid) at four different wavelengths, with fits (dashed).
    \textit{Inset:} rate parameter $a$ against rest-frame wavelength (blue), with cubic fit (orange), and the fit for the individual wavelengths plotted in the main panel (coloured crosses).
    }
    \label{fig:dimming_distribution}
\end{figure}

We first use our sample of simulated galaxies to obtain an estimate of the distribution of the `orientation factor' as a function of rest--frame wavelength.
By sampling from the distribution of `orientation factor' for all galaxies and orientations combined, we fold in the effect of both orientation and morphology simultaneously.
For each orientation $i \in [n]$ of each simulated galaxy $g$, where $n = \{1, \cdots, 50\}$, we calculated the normalised emission at wavelength $\lambda$,
\begin{equation}
  D^{i}_{g} = 1 - \frac{S^{i}_{g}(\lambda)}{\max\limits_{i \in n} \, S^{i}_{g}(\lambda)} \;\;,
\end{equation}
where $D$ can be considered the orientation factor.
\fig{dimming_distribution} shows the distribution of $D$ for all galaxies and orientations combined, at seven different wavelengths.
This distribution can be approximated by
\begin{equation}
  D = \frac{a\,e^{-ax}}{b} \;\;,
\end{equation}
where $a$ is analogous to the rate parameter in the exponential distribution, and $b$ sets the normalisation.
We fit to the distribution for a large range of rest--frame wavelengths (20--1000 \mumetre) and then fit the evolution of $a$ with $\lambda_{\mathrm{rest}}$ with a third--order polynomial, shown inset on \fig{dimming_distribution}.\footnote{coefficients: $c_1 = 2.4 \times 10^{-10}$, $c_2 = -5.6 \times 10^{-6}$, $c_3 = 6.8 \times 10^{-3}$, $c_4 = 5.3 \times 10^{-1}$}
We can now, for an \textit{arbitrary} rest--frame wavelength, sample from the orientation factor distribution by sampling from an exponential with corresponding rate parameter $a$.

\begin{figure}
	\includegraphics[width=\columnwidth]{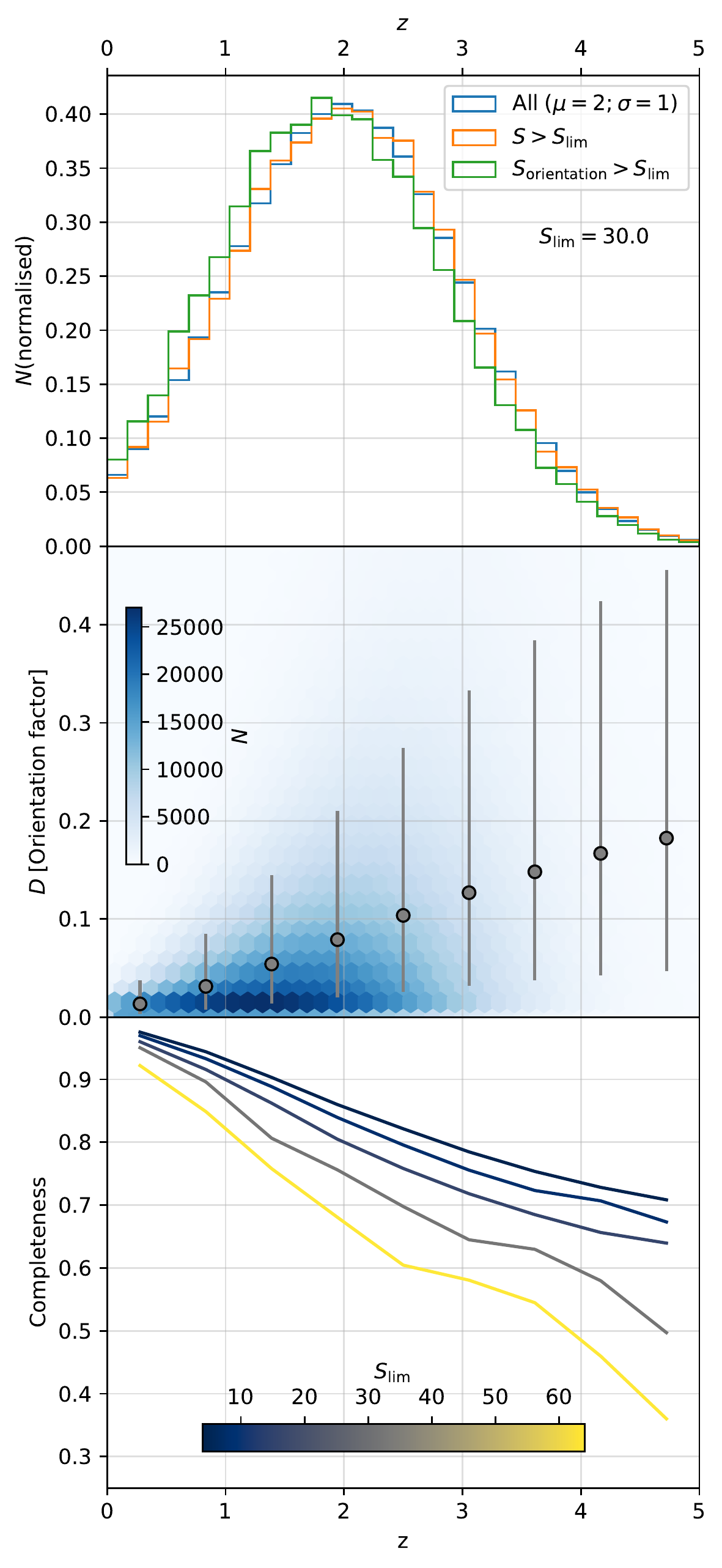}
    \caption{
      Redshift distribution of sources in a mock 250 \mumetre\ survey.
      \textit{Top:} The normalised redshift distribution of all sources (blue), those whose original flux density is above $S_{\mathrm{lim}} = 30$ (orange), and those whose flux density after including orientation effects is above $S_{\mathrm{lim}}$. There is a slight bias towards lower redshifts as a result of orientation effects.
      \textit{Middle:} the distribution of the orientation factor for the mock galaxies.
      Grey points show the median distribution binned in redshift, with errors showing the 16$^{\mathrm{th}}$--84$^{\mathrm{th}}$ percentile range.
      The 2D histogram (blue) shows the distribution of all sources.
      Galaxies at higher redshift are more likely to experience high orientation effects.
      \textit{Bottom:} completeness as a function of redshift $z$ at a given $S_{\mathrm{lim}}$.
      The redshift-dependent orientation factor leads to a redshift-dependent incompleteness.
    }
    \label{fig:redshift_completeness}
\end{figure}

The final two components of the model, the number density and redshift distributions, can be adjusted to accommodate any survey at a chosen wavelength.
Below we demonstrate an application of the model to a survey at $\lambda_{\mathrm{obs}} = 250$ \mumetre.
We begin by sampling from a \cite{schechter_analytic_1976} function for the flux density distribution,
\begin{equation}
  \frac{\mathrm{d}N}{\mathrm{d \,log}S} = \mathrm{log(10)} \; \frac{1}{S_{0}} \left( \frac{S}{S_{0}} \right)^{1-\gamma} \mathrm{exp} \left( - \frac{S}{S_{0}} \right) \;\;,
\end{equation}
where $S_{0} = 31.6 \; \mathrm{mJy}$ and $\gamma = 1.91$ \citep[fit to the counts from][]{wang_multi-wavelength_2019}.
Assuming a survey area of $10 \, \mathrm{deg^{2}}$, we sample galaxies above $S \geqslant 1$\,mJy; producing approximately $3\,900\,000$ galaxies, where the brightest has a flux density of $S\sim 260 \; \mathrm{mJy}$.
The redshift distribution is modelled using a simple truncated gaussian \citep[$\mu = 2$; $\sigma = 1$;][]{dunlop_blast_2010,mitchell-wynne_hermes_2012}, clipped between $z = [0,5]$, shown in the top panel of \fig{redshift_completeness}.

\begin{figure}
	\includegraphics[width=\columnwidth]{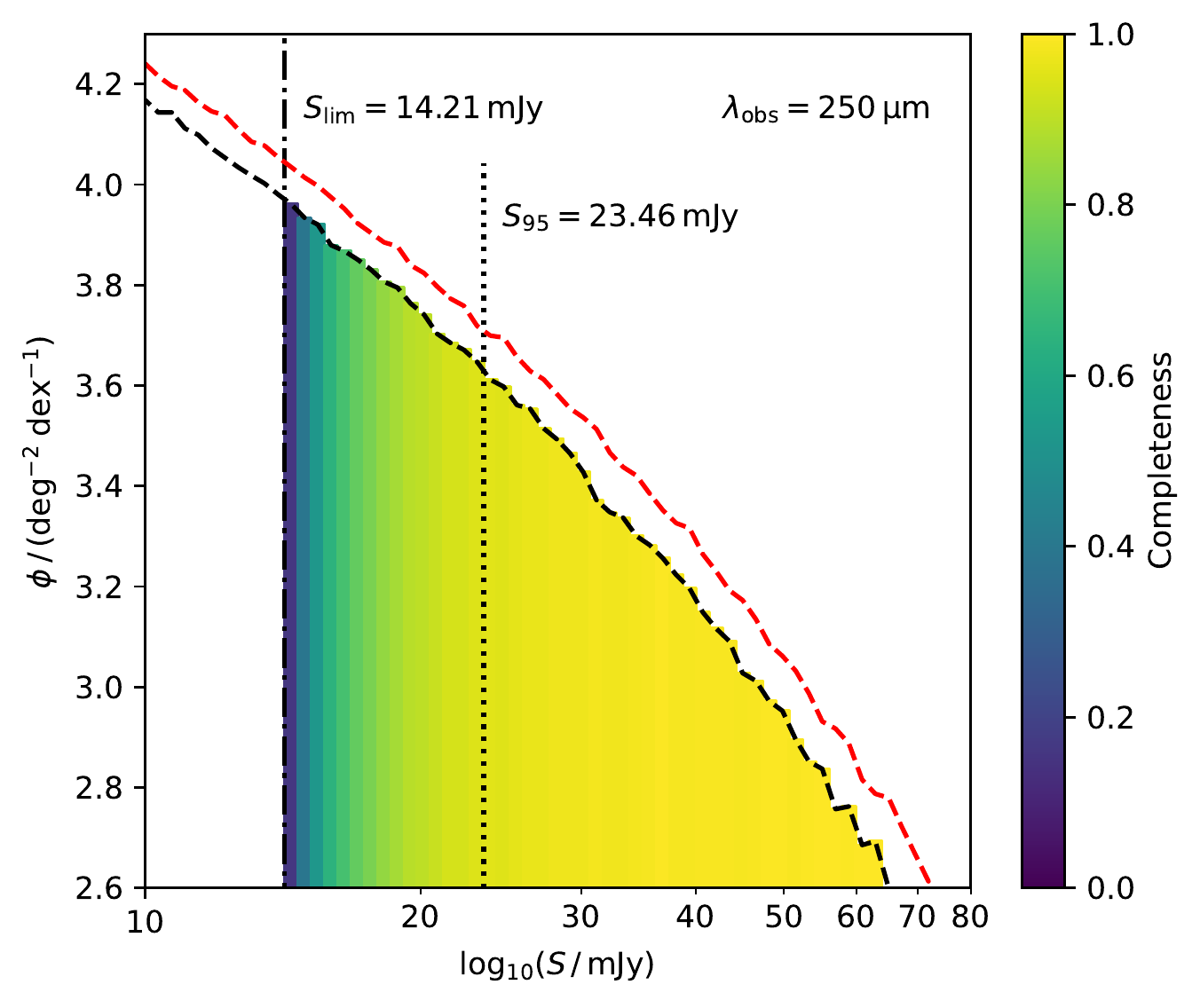}
    \caption{
    The observed $S_{250 \mathrm{\mu m}}$ number count distribution from our model.
    The red dashed line shows the number counts before the orientation factor is applied, the black dashed line shows the counts after orientation effects are considered.
    The counts including orientation effects are coloured by the orientation completeness, for a flux-density limit $S_{\mathrm{lim}} = 14.21 \, \mathrm{mJy}$ (dashed-dotted line).
    The flux density at which the completeness is 95\% ($S_{\mathrm{lim}} = 23.46 \, \mathrm{mJy}$) is shown by the dotted line.
    The completeness at a given flux density is dependent on both $S_{\mathrm{lim}}$ and wavelength.
    }
    \label{fig:completeness}
\end{figure}

The flux density for each randomly sampled galaxy $g$ at redshift $z$ is then
\begin{equation}
  S^{\,\mathrm{ori}}_{g} (\lambda_{\mathrm{obs}}) = S_{g} (\lambda_{\mathrm{obs}}) \, D_{g} (\lambda_{\mathrm{obs}}, z) \;\;.
\end{equation}
where the orientation factor is sampled from the previously estimated distribution at $\lambda_{\mathrm{rest}} = \lambda_{\mathrm{obs}} / (1+z)$, taking the corresponding value of $a$.
\fig{completeness} shows the number counts of $S_{250 \mathrm{\mu m}}$ sources both before and after including orientation effects, which reduces the normalisation of the abundance by $\sim$0.08\,dex.

\begin{figure}
	\includegraphics[width=\columnwidth]{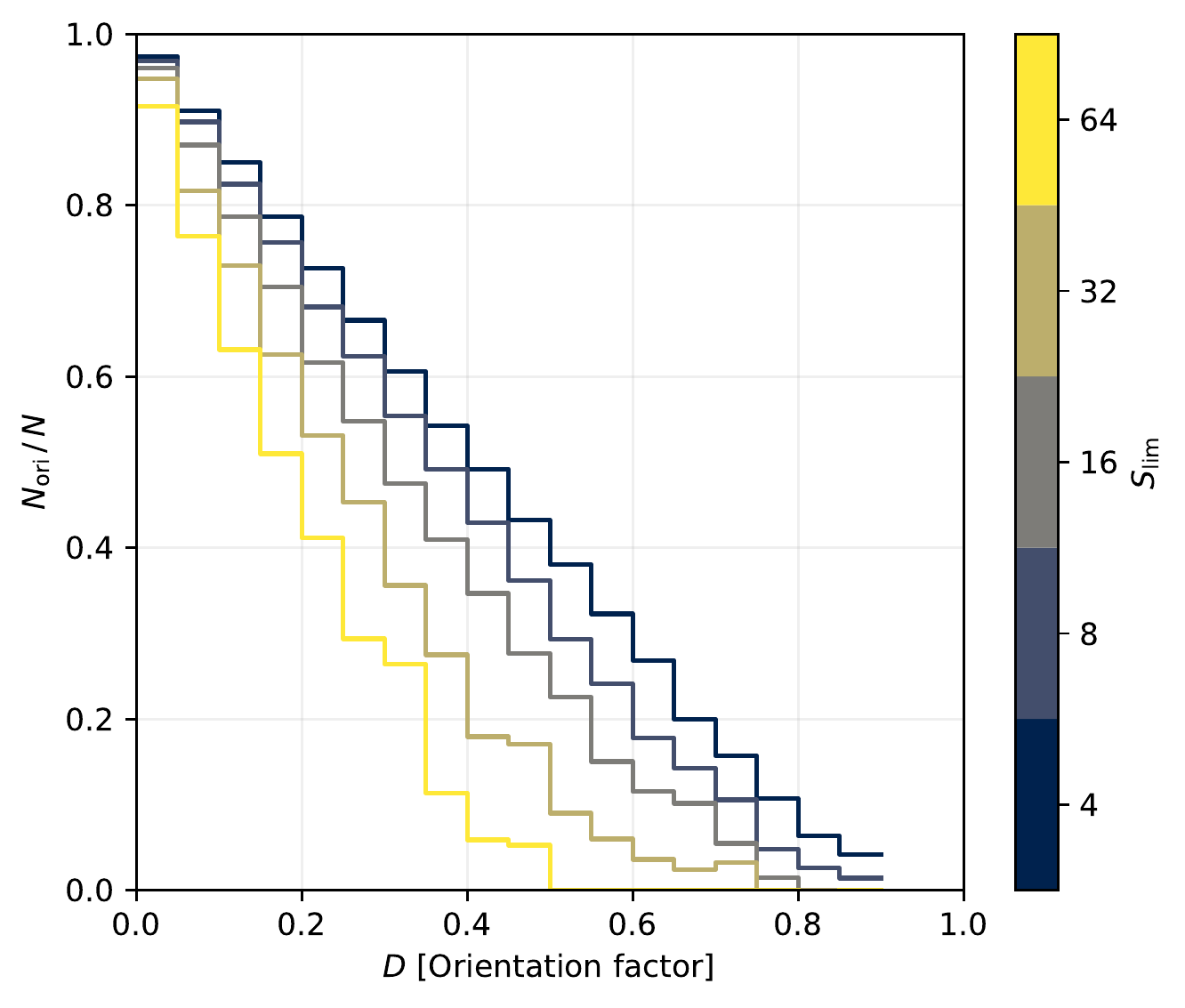}
    \caption{
      Fraction of galaxies recovered, binned by orientation factor $D$, at $\lambda_{\mathrm{obs}} = 250$ \mumetre.
      Each coloured line shows a different flux density limit ($S_{\mathrm{lim}}$).
			As expected, galaxies experiencing the greatest orientation effects (due to a combination of morphology and orientation) are less likely to be detected, and this effect is strongest for higher $S_{\mathrm{lim}}$.
    }
    \label{fig:dimming_fraction}
\end{figure}

We can now ask, given some flux-limited survey at a wavelength $\lambda$, how many galaxies of a particular orientation (encoded by $D$) are missed?
We can do this by comparing the number of galaxies in our survey including orientation effects, $N_{\mathrm{ori}}$, with those in the original survey, $N$.
\fig{dimming_fraction} shows the ratio $N_{\mathrm{ori}} \,/\, N$, which we refer to as the \textit{orientation completeness}, binned by the relative orientation factor $D$, for different survey limits at $\lambda = 250$\mumetre.
Where orientation effects are small, $D < 0.1$, at least 78\% of galaxies are recovered, but this falls dramatically for galaxies with increased orientation factor, down to $\sim$10\% for the most heavily obscured galaxies ($D > 0.4$) and a high flux density limit ($S_{\mathrm{lim}} = 64 \, \mathrm{mJy}$).

The number of galaxies missed with high orientation factors can be significant.
However, we expect the missed galaxies to be dominated by those close to the lower flux density limit, since (1) the orientation factor is applied equally to all galaxies, and (2) the abundance of galaxies is higher at lower flux densities.
We again use the concept of \textit{orientation completeness}, this time defined as the percentage of the original orientation distribution recovered, for a given lower flux density limit $S_{\mathrm{lim}}$, at a given flux density.
The top panel of \fig{completeness} shows the 250 \mumetre\ number counts, coloured by the completeness.
It is clear that at the bright end the completeness is close to 100\%, but this reduces at lower flux densities closer to $S_{\mathrm{lim}}$.
In this realisation, the flux density at which the completeness reaches 95\% is $S_{95} = 23.46 \, \mathrm{mJy}$, $9.25 \, \mathrm{mJy}$ greater than $S_{\mathrm{lim}}$.

Finally, since we model the redshift distribution of sources we can assess the completeness as a function of redshift.
The top panel of \fig{redshift_completeness} shows the normalised redshift distribution of all our modelled sources, as well as those above some flux density limit $S_{\mathrm{lim}}  = 30 \, \mathrm{mJy}$.
The distribution of galaxies above this limit after including orientation effects is slightly shifted to lower redshifts ($\Delta \, z_{\mathrm{median}} = -0.11$).
This can be explained by the evolution of the orientation factor distribution as a function of redshift, shown in the second panel of \fig{redshift_completeness}; at higher redshifts galaxies tend to have higher orientation factors, with a median at $z = 4.7$ of 11\%, compared to 2\% at $z = 0.3$.
This is a result of the peak of the orientation variability, shown in \fig{sigma_sed}, moving through the observed wavelength range.
Finally, the bottom panel of \fig{redshift_completeness} shows the completeness (defined as the fraction recovered: $N_{\mathrm{ori}} \,/\, N$) as a function of redshift for different $S_{\mathrm{lim}}$ cuts.
For $S_{\mathrm{lim}} = 64 \, \mathrm{mJy}$, the completeness is below 60\% at $z > 3$, which highlights the increasing impact of orientation incompleteness at high-$z$.

We have taken account of the effect of both orientation and morphology by folding them into a single `orientation factor' parameter.
In reality, these are two separate effects, and a more sophisticated model would independently treat the morphology of each object and its relative orientation.
However, this would require a much larger sample of simulated galaxies to fully sample the distribution of morphologies, and the subsequent RT to assess the effect on the orientation factor, which is computationally expensive.
It is nevertheless worth keeping in mind that the morphology has a known dependence on other physical properties, such as the stellar mass, and so the inferred distribution from our very small sample may not extrapolate to, for example, lower mass, sub-mm --faint galaxies.
The morphology distribution is also redshift dependent, which our fixed redshift sample does not take into account.

The analysis presented above at 250 \mumetre\ can be repeated for any survey at arbitrary wavelength, using our publicly available tool (\href{https://github.com/christopherlovell/orientation_bias}{github.com/christopherlovell/orientation\_bias}), and used to assess the impact of orientation bias for current or future surveys and instruments.
All of the analysis in this section is also agnostic to the chosen distribution of orientation factor, and can be repeated for arbitrary forms of this distribution motivated by observational or other theoretical work.

\section{Conclusions}
\label{sec:conc}

We have presented a prediction, based on dust radiative transfer modelling in hydrodynamic simulations, that submillimetre galaxy emission is dependent on orientation, and that this will lead to an \textit{orientation bias} in both the selection of sub-mm galaxies, as well as measurements of their properties.

Our findings in detail are as follows:
\begin{itemize}
  \item Sub-mm emission is orientation dependent, peaking at $\lambda_{\mathrm{rest}} \sim 55$ \mumetre\ (rest-frame), with variation of up to a factor of 2.7 at the peak of the emission between the different orientations.
  \item In ordered disc galaxies, the flux dependence can be parametrised by the relative orientation to the disc, with face-on galaxies showing higher emission than edge-on.
  However, we find no simple parameterisation for the orientation effect for arbitrary morphologies.
  \item This orientation dependence leads to a dispersion in properties measured from the sub-mm emission, such as the dust temperature and FIR luminosity (inter-percentile ranges of $5.0$ K and 0.110 dex, respectively), which translates into an uncertainty in the star formation rate of 30\% (when estimated using simple calibrations).
  \item The orientation dependence will also lead to an \textit{orientation bias} in the selection of galaxies in flux-limited samples.
  We model this bias, and find that the fraction of edge-on galaxies can be reduced by up to 80\% for a 250 \mumetre\ survey with a lower flux density limit of $S_{\mathrm{lim}} = 64 \, \mathrm{mJy}$.
  \item The orientation bias is also redshift dependent, leading to completeness in selected edge--on galaxies lower than 60\% at $z > 3$ for a lower flux density limit $S_{\mathrm{lim}} = 64 \, \mathrm{mJy}$ in a 250 \mumetre\ survey.
\end{itemize}

Together, these results suggest that the orientation bias will affect a number of surveys, as well as the statistical interpretation of targeted follow up studies.

We see the greatest variation in flux-density due to orientation close to the peak of the dust emission at $z = 2$ in the observer--frame.
At these wavelengths it is primarily \textit{Herschel}, with its PACS/SPIRE instruments, that has performed the deepest and widest surveys so far.
Unfortunately these instruments suffer from a high confusion noise level.
It is therefore unlikely that the orientation-bias we have seen in \simba\ will be detectable in such surveys, since they are limited to the most IR-luminous galaxies, or those sources suffering from significant multiplicity, even in those that are reasonably well cross-matched with optical surveys.
From the other perspective, ALMA's small field of view limits what can be achieved with blind surveys; those surveys that have been carried out \citep[\textit{e.g.}][]{franco_goods-alma_2018} are typically at longer wavelengths, where the orientation bias effect is lower.
However, ALMA is uniquely capable of resolved follow up of interesting sources, allowing us to understand the physics behind the orientation dependence in greater detail.
Follow up studies with ALMA will also be subject to the orientation bias effect, depending on the wavelength and instrument used to perform the initial selection.

Future planned, large single dish submillimetre observatories with much lower confusion limits will be more sensitive to fainter high-{\it z} SMGs selected at $\lambda_{\rm obs}>350$ \mumetre. For example,
the Atacama Large Aperture Submillimeter Telescope (AtLAST), currently in the design study phase, will allow high-resolution large-scale surveys at 350 \mumetre\ -- 4 $\mathrm{mm}$ thanks to its 50\,m single dish \citep{kawabe_new_2016,groppi_first_2019,klaassen_atacama_2020}.
These surveys will provide a direct means of testing whether the orientation bias we see in \textsc{Simba} is present, at least for $z > 2$.
If confirmed, we argue that this bias should be taken into account in future surveys with such instruments, particularly those used as source catalogues for detailed follow up with interferometric observatories such as ALMA -- these catalogues may be biased to galaxies with face-on orientations.
This uncertainty should also be taken into account when measuring and comparing the temperature and FIR luminosity distribution of sub-mm samples.

The code for generating models for user-specified surveys is publicly available at \href{https://github.com/christopherlovell/orientation_bias}{github.com/christopherlovell/orientation\_bias}.
It allows observers to assess the impact of orientation bias on a given survey, providing insight into the completeness in terms of flux density, redshift and orientation / morphology, for a given observed wavelength and combined (arbitrary) number count and redshift distribution.
We hope that this will be useful for understanding and assessing the impact of the orientation bias for past, present and future surveys.

\section*{Acknowledgements}
We wish to thank Jo Ramasawmy for help understanding the anticipated capabilities of AtLAST, and Aswin Vijayan for helpful discussions.
We acknowledge the following open source software packages used in the analysis (and not already referenced in the text): \textsf{scipy} \citep{2020SciPy-NMeth}, \textsf{Astropy} \citep{robitaille_astropy:_2013} \& \textsf{matplotlib} \citep{Hunter:2007}.
CCL acknowledges support from the Royal Society under grant RGF/EA/181016.
JEG is supported by a Royal Society University Research Fellowship.
MF acknowledges the support from STFC (grant number ST/R000905/1).
\simba\ was run on the DiRAC@Durham facility managed by the Institute for Computational Cosmology on behalf of the STFC DiRAC HPC Facility. The equipment was funded by BEIS capital funding via STFC capital grants ST/P002293/1, ST/R002371/1 and ST/S002502/1, Durham University and STFC operations grant ST/R000832/1. DiRAC is part of the National e-Infrastructure.

\section*{Data Availability}
The data underlying this article will be shared on reasonable request to the corresponding author.
The code to build and analyse the Monte Carlo model in \sec{bias} is available at \href{https://github.com/christopherlovell/orientation_bias}{\tt github.com/christopherlovell/orientation\_bias}.
The \simba\ simulation data and galaxy catalogs are publicly available at  \href{https://simba.roe.ac.uk}{\tt
https://simba.roe.ac.uk}.

\bibliographystyle{mnras}
\bibliography{smg_orientation,extra}

\appendix

\bsp	
\label{lastpage}
\end{document}